\documentstyle[eqsecnum,floats,preprint,aps,prd,epsf,rotate]{revtex}
\preprint{DAMTP-1999-101}
\date{5 August 1999}
\draft
\tighten
\begin{document}

\renewcommand{\arraystretch}{1.5}
\newcommand{\be}{\begin{equation}}
\newcommand{\ee}{\end{equation}}
\newcommand{\bea}{\begin{eqnarray}}
\newcommand{\eea}{\end{eqnarray}}
\def\Tr{\mathop{\rm Tr}\nolimits}
\def\sqr#1#2{{\vcenter{\hrule height.3pt
      \hbox{\vrule width.3pt height#2pt  \kern#1pt
         \vrule width.3pt}  \hrule height.3pt}}}
\def\square{\mathchoice{\sqr67\,}{\sqr67\,}\sqr{3}{3.5}\sqr{3}{3.5}}
\def\today{\ifcase\month\or
  January\or February\or March\or April\or May\or June\or July\or
  August\or September\or October\or November\or December\fi
  \space\number\day, \number\year}

\def\Bbb{\bf}
\def\be{\begin{equation}}
\def\ee{\end{equation}}

\title{Non-BPS D8-branes and Dynamic Domain Walls 
in Massive IIA Supergravities}

\author{A. Chamblin\thanks{H.A.Chamblin@damtp.cam.ac.uk},
M.J. Perry\thanks{malcolm@damtp.cam.ac.uk} and
H.S. Reall\thanks{H.S.Reall@damtp.cam.ac.uk}}

\address {\qquad \\University of Cambridge\\ DAMTP\\
Silver Street\\
Cambridge, CB3 9EW \\United Kingdom}

\maketitle

\begin{abstract}

We study the
D8-branes of the Romans massive IIA supergravity theory using the
coupled supergravity and worldvolume actions. D8 branes
can be regarded as domain walls with the jump in the extrinsic
curvature at the brane given by the Israel matching conditions. We
examine the restrictions that these conditions place on extreme and
non-extreme solutions and find that they rule out some of the
supersymmetric solutions given by Bergshoeff {\em et al}.
We consider what happens when the dilaton varies on the worldvolume
of the brane, which implies that the brane is no longer static.  
We obtain a family of D8-brane solutions parametrized by a
non-extremality term on each side of the brane and the asymptotic
values of the 10-form field. The non-extremality parameters can be
related to the velocity of the brane.
We also study 8-brane solutions of a  massive IIA supergravity theory
introduced by Howe, Lambert and West.  This theory
also admits a 10-form formulation, but the 10-form is not a R-R sector
field and so these 8-branes are not D-branes.  

\end{abstract}


\section{Introduction}

Perhaps the most important thing which string duality teaches us is 
that in order
to have a consistent string theory we have to include objects known as
``D-branes''.  These D-branes, which are just hyperplanes where open strings
are allowed to end, were shown by Polchinski \cite{joe} to be the carriers
of ten dimensional Ramond-Ramond (R-R) charge.  
T-duality requires the existence
of D-branes \cite{dai} and, once stringy effects are taken into account,
the D-branes become dynamical objects.  Indeed, the collective coordinates
for the transverse fluctuations of a D-brane are just the massless open string
excitations that live on the brane worldvolume.

Once one recognizes the existence of D-branes,
it is natural to start asking questions about their gross kinematical and/or
dynamical properties.  For example, is there any constraint on the topology of
these branes?  What happens when we push a brane away from extremality?
Can these branes rip, or tear, by some semi-classical decay process
analagous to the decay of cosmic strings (by black hole pair creation)?

Before proceeding with the principal construction, it is useful if we first
recall some basic facts about D-branes \cite{larus}.  To begin, let us focus
just on the bosonic terms in the effective field 
theory actions which arise from the IIA
and IIB string theories.  These field theories are of course just the
type IIA and type IIB supergravity theories.  
In the Neveu-Schwarz-Neveu-Schwarz(NS-NS)
sector these two theories have identical field content, consisting
of a metric tensor $g_{{i}{j}}$, an antisymmetric rank two tensor
potential $b_{{i}{j}}$ and a scalar dilaton field $\phi$.  
On the other hand, in the Ramond-Ramond (R-R) sector
the field content of the two theories is quite different.  Explicitly, the
bosonic sector for the effective action of IIA supergravity is given
as\footnote{We use a positive signature metric and the curvature
convention for which de Sitter space has positive Ricci scalar.}
\begin{eqnarray}
S_{IIA} &=&  {1\over 2\kappa^2} \int d^{10}x \, \sqrt{-g}
\Bigl[e^{-2\phi}
\bigl(R+4(\partial\phi)^2 \nonumber \\
&  & -{1\over 2\cdot 3!}H^2 \bigr)  
-\bigl({1\over 2{\cdot}2!}{F_{(2)}}^2 
+ {1\over 2{\cdot}4!}{F_{(4)}}^2 \bigr) \Bigr] \nonumber \\
&  & -{1\over 4\kappa^2} \int F_{(4)}\wedge F_{(4)}\wedge b  \>,
\end{eqnarray}
whereas the effective action for IIB supergravity takes the form
\begin{eqnarray}
S_{IIB} &=&  {1\over 2\kappa^2} \int d^{10}x \, \sqrt{-g}
\Bigl[e^{-2\phi}
\bigl(R+4(\partial\phi)^2 \nonumber \\
&  & -{1\over 2\cdot 3!}H^2 \bigr)  
-{1\over 2}(\partial A_{(0)})^2 \nonumber \\
&  & -{1\over 2\cdot 3!}(A_{(0)}B+F_{(3)})^2 \Bigr] \>, 
\end{eqnarray}
(where we have omitted terms involving the self-dual five-form which cannot
be expressed in a covariant fashion).
Here, $F_{(n)} = ndA_{(n-1)}$ are the antisymmetric 
$n$-form R-R field strengths, with
$(n-1)$-form potentials $A_{(n-1)}$, and $H = 3db$ is the antisymmetric
NS-NS three-form field strength.  The string coupling $g_s$ is 
given in terms of
the dilaton by the relation $g_s = e^{\phi}$.  Thus, in both of the above
actions the NS-NS sector is multiplied by a factor of ${g_s}^{-2}$; this means 
that NS-NS effects arise in ordinary string perturbation theory.  On the other
hand, the R-R terms are not multiplied by any such factor and so we know that
fundamental string states do not carry R-R gauge charges.  As we have already 
mentioned, the mystery of the R-R gauge fields was resolved when it was shown
that R-R charge was carried by the topological defects 
known as D-branes \cite{joe}.

These actions do not tell the whole story about D-brane configurations.
This is because they really only tell us about what life is like for massless
closed string states which exist ``off'' the D-branes and how these states
couple to the D-branes.  For this reason, the supergravity effective actions
are called the ``bulk'' terms.  
There should also be an effective description of 
life on the brane worldvolume.  Indeed, such an effective world-volume action
(denoted $S_{WV}$) does exist and it is characterized by a D(p)-brane tension,
${\mu}_{p}$, and R-R charge density form $f_{p}$ as shown below:
\begin{equation}
\label{eqn:wvaction}
S_{WV} = -{\mu}_p {\int} d^{p+1} \sigma e^{- {\phi}} 
\sqrt{\det (h_{\mu \nu} + b_{\mu \nu} + 2 \pi {{\alpha}^{\prime}}
F_{\mu \nu})} - \nonumber \\
f_{(p)} {\int} d^{p+1} \sigma A_{(p+1)}
\end{equation}
Here, $\sigma$ denotes the coordinates tangent to the brane 
worldvolume, $h_{{\mu}{\nu}}$, $b_{{\mu}{\nu}}$ and $\phi$ are the 
pullback of the (respective) spacetime fields to the brane worldvolume.
$F_{\mu \nu}$ is an ordinary Maxwell gauge field which lives on the brane
worldvolume.  Again, $A_{(p+1)}$ is an antisymmetric $(p+1)$-form R-R potential
which couples naturally to the p-brane worldvolume (hence the appearance of
the R-R charge density $f_{(p)}$).  This form of the action is very useful
if one is interested in studying ``light'' branes, which have negligible
gravitational fields (so that one can ignore the bulk supergravity 
contributions). 
In this limit a beautiful picture emerges, in which fundamental
strings ending on a D-brane appear, from the point of view of the
brane worldvolume theory, as Coulomb-like point particle solutions
of the Born-Infeld theory on the brane (Gibbons refers to all such solitonic
configurations in Born-Infeld theory as ``BIons'') \cite{gazz,curt}.  
Similarly, one can also think of
an M2 brane ending on an M5 brane as a ``vortex''-type BIon living on the
M5 brane worldvolume, and so on.  While these are certainly beautiful and
promising results, it is still important to understand how strings and
branes in general will interact when they are heavy i.e. the effects of
gravity are taken into account.  Presumably, addressing this problem will have
to involve somehow writing down an effective action which interpolates between
the world-volume and bulk terms as gravity is turned up.
In this paper, we consider the simpler problem of how the
world-volume and bulk terms interact with each other when the
branes gravitate. 

We concentrate on the case of D8-branes of massive IIA supergravity
because we can then use the theory developed in \cite{great} to solve
the equations of motion of the coupled worldvolume-bulk system for
moving branes. D8-branes can be viewed as domain walls,
which have been extensively
studied in cosmology and supergravity. The gravitational effects of a
domain wall are described by the Israel equations which are discussed
in section \ref{sec:israel}. The massive IIA supergravity theory and
its D8 branes are described in \ref{sec:romans}. In section
\ref{sec:dynamic} we give our solutions describing moving D8
branes. Section \ref{sec:hlw} describes dynamic branes of a different
massive IIA supergravity and we conclude in section \ref{sec:conclude}.

\section{The Israel matching conditions and dynamic dilatonic domain walls} 

\label{sec:israel}

The fact that 8-branes divide ten-dimensional spacetime up into
domains is reminiscent of domain walls in cosmology.
As is well known, various cosmological models assume that a variety of
phase transitions took place in the early universe.  In such transitions,
symmetries which are only valid at high temperatures are broken as the
universe cools down.  Bubbles of the new phase
are nucleated in regions of the old phase and may expand; if the rate of
production of these bubbles is not diluted by the rate of expansion of the
universe, the process of bubble nucleation, expansion and amalgamation
will continue until the universe is filled with new phase (with perhaps a few
bubbles of the old phase left over).  When the new phase fills the
entire universe, the transition is said to be complete.

The dynamical evolution of these bubbles  
when the effects of gravity are included, has been studied by a number of
authors \cite{ber,guth}.  
These studies involve understanding
the Einstein equations when the source is a thin shell, or domain wall.
In such situations, the spacetime has low differentiability and one
has to regard the curvature as a distribution.
It was shown long ago \cite{israel} that the correct formalism for studying
such a problem involves constructing metric junction conditions for the
thin shells.  These junction conditions, commonly referred to as the
``Israel matching conditions'', state the discontinuity in the
extrinsic curvature of the shell is related to the energy-momentum
tensor $t_{ab}$ of the matter on the shell by
\be
\label{eqn:israel}
 [K_{ab}-Kh_{ab}]=8\pi G t_{ab},
\ee
where $K$ denotes the trace of the extrinsic curvature $K_{ab}$ and 
$h_{ab}$ is the induced metric on the shell. A simple derivation of
this equation is given in \cite{great}.
Thus the energy momentum of an idealized domain wall gives rise to a 
jump in the 
extrinsic curvature of surfaces parallel to the wall as one moves through 
the wall.
These conditions are easily satisfied
when the only bulk energy density comes from a cosmological constant
and the energy density of the wall is constant. In supergravity
theories one typically has scalar fields present that couple to domain
walls and until recently solutions had only been found when these
fields are constant over the worldvolume of the wall \cite{cvetic,cvetic2,mir}.
A method for dealing with the case of non-constant worldvolume fields
was given in \cite{great}. If one assumes that the bulk on both sides
of the wall is \emph{static} then consistency of the Israel 
conditions yields non-trivial relationships between the metric in the
bulk and the matter on the wall. These can be used to solve the Einstein
equations in the bulk. The resulting spacetime usually contains
cosmological horizons so, although one starts from a static ansatz,
the spacetime is actually time-dependent.
We will use these results when we describe the dynamic 
branes of the Romans theory because there the dilaton has a 
Liouville potential and therefore has to run in the bulk.

\section{D8 branes and the Romans theory}

\label{sec:romans}

The D8-brane of IIA string theory couples to the ten-form field strength
$F_{10}$ of the R-R sector.  As Polchinski \cite{joe} pointed out, this
ten-form is not a dynamical variable; rather, it is just a constant field
which generates a uniform physical energy density which permeates space.
This energy density is proportional to the square of the mass
term of the massive IIA supergravity theory derived by Romans \cite{roman}.  
In this theory, the mass
arises from a Higgs mechanism in which the two-form ``eats'' the vector.  

The ten form formulation of the Romans massive IIA supergravity was given by
Bergshoeff {\em et al}. \cite{bergshoeff}. For our purposes the
relevant part of the action in the string frame is
\be
 S = \frac{1}{2{\kappa}^2} \int d^{10}x \sqrt{-g} \left(e^{-2\phi} (R +
4(\partial \phi)^2) -\frac{1}{2} M^2 \right) +
\frac{1}{2\kappa^2}\int \frac{1}{10} M F_{(10)},
\ee 
where $M$ is an auxiliary field which can be eliminated via its
equation of motion. The Einstein frame metric is given by
$g_{ab}^{(E)} = e^{-\phi/2}g_{ab}^{(S)}$. In this frame the action is
\be
 S = \frac{1}{2{\kappa}^2} \int d^{10}x \sqrt{-g}
(R-\frac{1}{2}(\partial \phi)^2-\frac{1}{2}e^{5\phi/2}M^2) +
\frac{1}{2\kappa^2} \int\frac{1}{10} M F_{(10)}.
\ee
We shall set $\kappa = 1$. 
The ten form is given in terms of its nine form
potential by $F_{(10)} = 10dA_{(9)}$. Varying $A_{(9)}$ gives
$M=\mathrm{const.}$ and varying $M$ gives $F_{(10)} = 10 M
e^{5\phi/2}\eta _{(10)}$ where $\eta_{(10)}$ is the volume form
of the spacetime. 
Note that varying the metric does not affect the final term. Therefore
the Einstein equations are the same as those for which the only matter
is a scalar field $\phi$ with potential 
\be
 V(\phi) = \frac{1}{4}M^2 e^{5\phi/2}.
\ee

\bigskip

The relevant part of the D8 brane string frame action is \cite{larus}
\be
 S_{brane} = -\mu\int d^9\xi \sqrt{-h} e^{-\phi} \pm \mu \int A_{(9)},
\ee
where 
$h$ is determinant of the induced metric on the brane world volume and the
final term involves the pull back of the bulk nine form
potential to the brane's world volume. The upper sign refers to a brane
and the lower sign to an anti-brane. In the Einstein frame the action is
\be
 S_{brane} = -\int d^9\xi \sqrt{-h} \hat{V}(\phi) \pm \mu \int
A_{(9)}
\ee
where
\be
\label{eqn:hatV}
 \hat{V}(\phi) = \mu e^{5\phi/4}.
\ee
Note that the final (Wess-Zumino) term 
is independent of the metric so it does not
contribute to the world volume energy momentum of the brane and
therefore does not occur in the Israel equations. However the brane is
a source for the ten form so we expect $F_{(10)}$ to have a
discontinuity at the brane arising from this term. To see this,
vary $A_{(9)}$ in the total action $S_{bulk}+S_{brane}$ to obtain the
surface term
\be 
 \frac{1}{2} \int ([M] \pm 2\mu)\delta A_{(9)},
\ee
where we refer to the bulk regions on the two sides of the brane as
$(+)$ and $(-)$ and $[M]\equiv [M]^{+}_{-}$ denotes the discontinuity of $M$ at
the brane\footnote{
The signs are determined as follows. 
Choose an oriented atlas for the whole spacetime
(including the brane) and consider a chart $x^a$ defined in a
neighbourhood of the brane such that the brane is at $x^9=0$. Let
$(+)$ denote the region $x^9<0$, $(-)$ the region $x^9>0$ and $n$
the outward unit normal to the $(+)$ region. The volume form of the bulk at the
brane is $\eta_{(9)}\wedge n$ where $\eta_{(9)}$ is the volume form of
the brane world volume.}.
If this term is to vanish then we need
\be
\label{eqn:Mjump}
 [M] =\mp 2\mu,
\ee
which relates the jump in the field strength to the charge of the brane.

There will also be a discontinuity in the normal derivative of the
dilaton at the brane. Varying the dilaton gives a surface term
\be
 -\int d^9\xi \sqrt{-h} \left(\frac{1}{2}[n.\partial
\phi]+\frac{d\hat{V}}{d\phi}\right) \delta\phi,
\ee
where $n$ is the unit normal pointing away from the $(+)$ region of
the bulk. 
The surface term vanishes for arbitrary $\delta\phi$ only if
\be
\label{eqn:phijump}
 [n.\partial\phi]=-2\frac{d\hat{V}}{d\phi}.
\ee
This equation ensures that energy-momentum is conserved as the brane
moves through regions of varying $\phi$ \cite{great}. 

\section{Dynamic D8-branes of the Romans theory}

\label{sec:dynamic}

We can apply the method developed in \cite{great} to solve the
equations of motion of the coupled brane-bulk system. The starting
point is the assumption that the metric is static on both sides of the
brane. It will turn out that this assumption is less restrictive 
than it may seem because the bulk spacetime typically contains
cosmological horizons beyond which it is manifestly time dependent.
We shall assume that on each side the bulk metric can be written
\be
\label{eqn:metricansatz}
 ds^2 = -U_{\pm}(r_{\pm})dt_{\pm}^2 + U_{\pm}(r_{\pm})^{-1} dr_{\pm}^2
+ R_{\pm}(r_{\pm})^2d\Omega_k^2,
\ee
where $d\Omega_k^2$ is the line element on an eight dimensional 
Einstein space with
metric $\bar{g}_{mn}$ and Ricci tensor $\bar{R}_{mn}=7k\bar{g}_{mn}$
with $k\in \{-1,0,1\}$. The dilaton is assumed to be a function only of
$r_{\pm}$. The bulk field equations (i.e. the Einstein equations and
the dilaton equation of motion) for this ansatz were given in
\cite{great}\footnote{The equations in \cite{great} use the rescaled
dilaton $\phi/\sqrt{2}$.}. 
Let the brane position be $t_{\pm}=t_{\pm}(\tau),r_{\pm}=r_{\pm}(\tau)$ where
$\tau$ denotes proper time of an observer at rest relative to the
spatial sections of the brane i.e. the induced metric on the brane
take the Friedmann-Robertson-Walker form
\be
 ds^2 = -d\tau^2 + R(\tau)d\Omega_k^2.
\ee
Continuity of the metric at the brane requires that
$R_+(r_+(\tau))=R_-(r_-(\tau))=R(\tau)$. Henceforth we shall
drop the $\pm$ subscripts except where this might cause confusion.

The proper velocity of the brane is
\be
 u^a = \left(\sigma U^{-1} E, \frac{dr}{d\tau},0,\ldots,0\right),
\ee
where $\sigma=+1$ when $U(r)>0$ i.e. when $t$ is a time coordinate. If
$t$ becomes a spatial coordinate (e.g. after crossing a horizon) then
$\sigma = \pm 1$. The quantity $E$ is the energy per unit mass of
a particle comoving with the brane
\be
 E = \sqrt{U + \left( \frac{dr}{d\tau} \right)^2}.
\ee
The unit normal is
\be
 n_a = \epsilon\left(\sigma\frac{dr}{d\tau},-U^{-1}E,0,\ldots,0\right),
\ee
where $\epsilon = \pm 1$ is chosen so that the
normal points \emph{out} of the $(+)$ region and \emph{into} the $(-)$
region. 

The extrinsic curvature $K_{ab} \equiv h_a^c h_b^d \nabla_c n_d$ 
is most easily computed in the basis $u^a ,
n^a , e_{(1)}^a , \ldots, e_{(8)}^a$, where the $e_{(i)}^a$ are a
basis for the spatial sections $t=\mathrm{const.}$,
$r=\mathrm{const.}$ See \cite{great} for the details of this
calculation. The result is
\be
\label{eqn:k00}
 K_{00} = \epsilon \frac{dE}{d\tau}/\frac{dr}{d\tau}, 
\ee
\be
 K_{ij} = -\epsilon E\frac{R'}{R}h_{ij},
\ee
where a prime denotes a derivative with respect to $r$. The Israel
equations \ref{eqn:israel}
relate the jump in the extrinsic curvature to the brane
world volume energy momentum tensor,
\be
 t_{ab} = -\hat{V}(\phi)h_{ab}.
\ee
They can be rewritten as
\be
 [K_{ab}] = \frac{\hat{V}(\phi)}{8} h_{ab},
\ee
so the $ij$ components reduce to
\be
\label{eqn:israelij}
 \left[\epsilon E\frac{R'}{R}\right] =
-\frac{1}{8}\hat{V}(\phi).
\ee
Note that $\epsilon_+$ and $\epsilon_-$ need not be the same. Equation
\ref{eqn:phijump} gives the discontinuity in the normal derivative of
$\phi$
\be
\label{eqn:phijump2}
 \left[\epsilon E \phi'\right] =
2\frac{d\hat{V}}{d\phi}.
\ee
The similarity of these two equations allows us to relate $\phi$ and
$R$. Note that $d\phi/dR$ must be continuous across the brane and that
\be
 \phi_{\pm}' = \frac{d\phi}{dR} R_{\pm}'.
\ee
This can be substituted into equation \ref{eqn:phijump2} and compared
with equations \ref{eqn:israelij} to yield
\be
\label{eqn:dphidR}
 \frac{d\phi}{dR}=-\frac{16}{R\hat{V}}\frac{d\hat{V}}{d\phi}.
\ee
This equation has to hold at every point visited by the brane so if the
brane moves through a range of $R$ then it can be integrated to yield $\phi$
as  function of $R$. 

We now turn to the $00$ component of the Israel conditions, which can
be written
\be
\label{eqn:k00a}
 \frac{d}{d\tau}\left[\epsilon E \frac{R'}{R}\right] -
\frac{1}{R}\frac{dR}{d\tau} \left[\epsilon E \frac{R'}{R} X\right] =
-\frac{\hat{V}}{8R}\frac{dR}{d\tau},
\ee
where
\be
 X \equiv \left(\frac{R'}{R}\right)' \left(\frac{R'}{R}\right)^{-2}.
\ee

In \cite{great} the bulk spacetime was assumed to be symmetric under
reflection in the brane. Consistency of the $00$ and $ij$ components
of the Israel equations was then shown to reduce to a condition
relating $R'$ and $\hat{V}$.
This argument does not appear
possible here without appealing to the bulk 
field equations\footnote{We would need
to show that $X$ is continuous across the brane, allowing us to take
it outside the square brackets in equation \ref{eqn:k00a} and then
use equation \ref{eqn:israelij} to eliminate the square
brackets. However it does not appear possible to show that $X$ must
take the same value on both sides of the brane without using the bulk
field equations}, which is what we shall now do. One of the Einstein
equations can be written \cite{great}
\be
 \frac{R''}{R} = -\frac{1}{16}{\phi'}^2.
\ee
Combining this with equation \ref{eqn:dphidR} and integrating gives
\be
\label{eqn:RV}
 R_{\pm}'(r_{\pm}) = C_{\pm}\hat{V}(\phi(r_{\pm})),
\ee
where $C_{\pm}$ are constants of integration. Equations
\ref{eqn:dphidR} and \ref{eqn:RV} relate the bulk metric and
dilaton. It is easily verified that these, together with equation
\ref{eqn:israelij} ensure that equation \ref{eqn:k00a} is satisfied.

Substituting in the explicit form for $\hat{V}$ given by
equation \ref{eqn:hatV} allows one to solve for
$R_{\pm}(r_{\pm})$ and $\phi_{\pm}(r_{\pm})$. These can then be
substituted into the bulk field equations to solve for
$U_{\pm}(r_{\pm})$, as described in \cite{great}. Solutions only exist
when $k=0$ (i.e. for Ricci flat spatial sections\footnote{
In the nomenclature of \cite{great} these would be described as type
II solutions.}). The bulk solution takes the
same form on both sides of the brane:
\be
 U_{\pm}(r_{\pm}) = 26^2 r_{\pm}^{1/13} \left(2^{-8} M_{\pm}^2 
e^{5\phi_{0\pm}/2} -
2m_{\pm} r_{\pm}^{8/13}\right),
\ee
\be
 R_{\pm}(r_{\pm}) = r_{\pm}^{1/26},
\ee
\be
 \phi_{\pm}(r_{\pm}) =  \phi_{0\pm} - \frac{10}{13} \log r_{\pm},
\ee
where $\phi_{0\pm}$ and $m_{\pm}$ are constants of integration.
Continuity of the metric at the brane requires
$R_+(r_+(\tau))=R_-(r_-(\tau))$ hence
$r_+(\tau)=r_-(\tau)$. Continuity of the dilaton at the brane then gives
$\phi_{0+} = \phi_{0-}$. We can now drop the $\pm$ subscripts when
referring to $R(r)$, $r(\tau)$ or $\phi(r)$. 

By rescaling the
coordinates it is possible to either set $\phi_0=0$ or
$|m_{\pm}|=1$ whilst preserving the form for the metric in equation 
\ref{eqn:metricansatz}, so the solutions form a 1-parameter family. 
However continuity of the metric of the spatial sections does not permit
independent rescalings on both sides of the brane so it is not in
general possible to set $|m_+|=|m_-|=1$. We shall rescale so that
$\phi_0=0$. We shall refer to the maximal
analytic extension of this solution as the maximal bulk solution to
distinguish it from the solution when the jump in extrinsic curvature
at the brane is taken account of.

The maximal bulk solution is singular at $r=0$. The singularity is
timelike. If $m_{\pm}$ is positive then there is a cosmological horizon
beyond which $r_{\pm}$ becomes a time coordinate and the metric
non-static. However, as discussed in \cite{great}, our solutions
remain valid when the brane crosses a horizon. If $m_{\pm}$ is
negative then the solution is static everywhere. When $m_{\pm}=0$ then
we recover the supersymmetric solution of Bergshoeff {\em et al}. 
\cite{bergshoeff} provided we take the 8 dimensional spatial
sections to be flat\footnote{More general supersymmetric solutions
have recently been discussed in \cite{dom}, where it was shown that some
supersymmetry will be preserved if the brane worldvolume is taken to
be any Ricci flat space admitting a Killing spinor.}.
We have therefore found a one-parameter family of
non-extremal generalizations of this supersymmetric solution. Note
that the asymptotic (large $r$) behaviour of the metric is determined
by the term involving $m_{\pm}$ so the non-extremal and supersymmetric
solutions have different asymptotic behaviour.
It is therefore not possible to define an energy for the spacetime relative
to the extreme solution by using a surface integral at infinity. However
$m_{\pm}$ is clearly analagous to a mass for the spacetime, so for some
of our discussion we shall assume that it is positive. If $M_{\pm}$ also
vanishes (so the 10-form vanishes) then our solution becomes
degenerate ($U=0$). However it is easy to see that in this case a
solution is $U=1$ with $\phi$ and $R$ constant. If the 8
dimensional spatial sections (with line element $d\Omega_0^2$) are
flat space then this solution is simply Minkowski space.

\bigskip

By squaring to eliminate the square
roots (in $E$) we obtain the brane's equation of motion
\be
\label{eqn:eqmotion}
 \frac{1}{2} \left( \frac{dR}{d\tau} \right)^2 + F(R) = 0,
\ee
where
\bea
 F(R) &=& -\frac{32}{\mu^2} [m]^2 R^{-16} - \left(\frac{1}{2}\{m\} \pm
\frac{1}{4\mu} [m] \{M\}\right) R^{-32} \nonumber \\
 &=& -\frac{32}{\mu^2} (m_+ - m_-)^2 R^{-16} \mp \frac{1}{2\mu}
(m_+M_- - m_- M_+) R^{-32}.
\eea
The curly brackets denote a sum over the sides of the wall
(i.e. $\{m\} = m_+ + m_-$)
and we have used equation \ref{eqn:Mjump} describing the discontinuity
in $M$ at the brane. The equation of motion is the same as for a
particle of unit mass and zero total energy moving in a potential
$F(R)$. The fact that the same coefficient $\mu$ occurs in the
Born-Infeld and Wess-Zumino terms of the D8-brane action makes
the coefficient of a possible $R^{-8}$ term in $F(R)$ vanish.
Specifying $r$, $dr/dt_+$ and $dr/dt_-$ at a given instant allows one
to determine $m_+$ and $m_-$ using this equation of motion. In this
sense, initial data determines the spacetime geometry.

Note that the only dependence of $F(R)$ on $M_{\pm}$ is through the
``interaction'' term $[m]\{M\}$. If $[m]=0$ then the brane does not
notice the jump in $M$. This is obvious on physical grounds since the
brane is the source for the jump in $M$ and will not move under the
influence of its own field. 

The equation of motion was obtained by squaring equation
\ref{eqn:israelij}. We
must now go back and check which solutions of the equation of motion
are still solutions of the unsquared equation. This can be written
\be
 \epsilon_+ |\mu^2-Y(R)| - \epsilon_- | \mu^2 + Y(R)| = -2\mu^2,
\ee
where
\be
 Y(R) \equiv 2^7[m]R^{16} \pm \frac{1}{2} \mu \{M\}.
\ee
(The upper sign is for a brane and the lower sign for an anti-brane.)
There are three possible solutions to this equation:
\bea
\label{eqn:epsilonchoices}
 Y(R) > \mu^2 &\Rightarrow & \epsilon_+ = +1, \qquad \epsilon_-=+1,
\nonumber \\
 |Y(R)|<\mu^2 &\Rightarrow & \epsilon_+ = -1, \qquad \epsilon_-=+1,
\nonumber \\
 Y(R) <-\mu^2 &\Rightarrow & \epsilon_+ =-1, \qquad \epsilon_-=-1.
\eea
Note that since $R$ can vary, it is possible for $\epsilon_{\pm}$ to
change sign as the brane moves. This can happen only when $E_{\pm}$ vanishes,
which is only possible if $U_{\pm}(r(\tau))$ is negative
i.e. $t_{\pm}$ is a spatial coordinate. Vanishing $E_{\pm}$
corresponds to $dt/d\tau=0$ which can happen if there is a change in
the direction of motion of the brane relative to the bulk spatial sections.
The geometrical interpretation of $\epsilon$ follows from
\be
 -n\cdot \partial R = \epsilon_{\pm} E_{\pm} R'.
\ee
The left hand side describes how $R$ varies as one approaches the
brane from the side into which $n$ points (i.e. the $(-)$ side) in a
direction perpendicular to the brane. Hence if $\epsilon_-=+1$ then
$R$ increases towards the brane on the $(-)$ side. If $\epsilon_-=-1$
then $R$ decreases towards the brane on the $(-)$ side. On the other
side, if $\epsilon_+=+1$ then $R$ decreases towards the brane on the
$(+)$ side (since $n$ points into the $(-)$ side) and if
$\epsilon_+=-1$ then $R$ increases towards the brane on the $(+)$
side. This information tells us which portions of the maximal 
bulk solution are
relevant on each side of the wall. For example if $\epsilon_-=+1$ then
we would take the maximal bulk solution, 
cut it along the brane at $r=r(\tau),
t=t(\tau)$ and keep the part for which $R$ increased towards the brane
in a perpendicular direction. This would give us the $(-)$ bulk
region. A similar construction would give the $(+)$ region and the two
are glued together at the brane.

Note that it is not possible to have $\epsilon_+=+1,\epsilon_-=-1$ which
corresponds to $R$ decreasing towards the brane in the perpendicular
direction on both sides. This would require a brane of negative
tension.

The following subsections discuss the behaviour of the brane in
several cases of interest.

\subsection{Static solutions}

The simplest case to consider is when the brane is static, by which we
mean $R(\tau)$ is constant. Our method does not necessarily yield all of the
solutions in this case since our bulk solutions were derived by
assuming that the brane swept
out a range of $r$ in the bulk. However if we assume that the bulk
solution on each side of the brane is part of the maximal solution
found above then we can
find static solutions.
Our analysis needs modifying in the
static case. The reason for this is the $dr/d\tau$ in the denominator
of the $00$ component of the extrinsic curvature, given by equation
\ref{eqn:k00}. If this term vanishes then we must use
\be
 K_{00} = \frac{\epsilon U'}{2\sqrt{U}},
\ee
which can be obtained from equation \ref{eqn:k00} by taking the limit
as $dr/d\tau \rightarrow 0$. Subtracting the $ij$ component of the
Israel equation from the $00$ equation yields
\be
 \left[\frac{\epsilon}{\sqrt{U}}\left(U'-2\frac{R'}{R}U\right)\right]=0.
\ee
Substituting in the expressions for $U_{\pm}(r)$ and $R(r)$ then leads to
\be
\label{eqn:statick00}
 \frac{\epsilon_+ m_+}{\sqrt{U_+}} = \frac{\epsilon_- m_-}{\sqrt{U_-}}.
\ee
The $ij$ component of the Israel equation becomes
\be
 \epsilon_+ \sqrt{U_+} - \epsilon_- \sqrt{U_-} = -\frac{13\mu R}{4}.
\ee
(We choose to work with this directly, rather than with the rearranged
form involving $F(R)$ derived above.)
These equations can be solved to give
\be
 \sqrt{U_{\pm}} = \mp\frac{13\epsilon_{\pm} \mu
R}{4(1-m_{\mp}/m_{\pm})},
\ee
from which it follows that $m_+ \ne m_-$. Assume that $m_+>m_-$. It
follows that $\epsilon_+=\epsilon_-=-1$. Solving for $R$ yields
\be
 R^{16} = \frac{1}{2^9m_{\pm}}\left(M_{\pm}^2 - \frac{4 m_{\pm}^2
\mu^2}{[m]^2} \right).
\ee
Consistency requires
\be
 \frac{1}{m_+}\left(M_+^2 - \frac{4m_+^2\mu^2}{[m]^2}\right) =
 \frac{1}{m_-}\left(M_-^2 - \frac{4 m_-^2 \mu^2}{[m]^2}\right).
\ee
Multiplying by $m_+ m_-/[m]$, setting $x=m_+/[m]$ and using
$[M]=\mp 2\mu$ gives
\be
 x^2 \pm \frac{M_+}{\mu} x + \frac{M_+^2}{4\mu^2}=0,
\ee
which has the unique solution $x=\mp M_+/(2\mu)$. This then yields
$R=0$, which is the singularity. This conclusion can only be avoided
if $m_+=m_-=0$. Therefore a static solution exists if,
and only if, the bulk on both sides of the brane is supersymmetric.

The supersymmetric maximal bulk solutions are globally static
with a timelike naked singularity at $r_{\pm}=0$.
It is straightforward to see that any static solution describing a
single D8-brane has to include these singularities. 
To avoid the singularities at
$r_{\pm}=0$, $r_+$ has to decrease towards the brane on the $(+)$ side
and $r_-$ has to decrease towards it on the $(-)$ side. This means
that $\epsilon_+ = +1$ and $\epsilon_- = -1$. However this is
the case that was ruled out above. Timelike singularities
in dilatonic domain wall spacetimes have been studied in \cite{cvetic2}.

Bergshoeff {\em et al}. gave supersymmetric configurations of
parallel branes with vanishing 10-form on both sides of the
``stack''. This is impossible once the
restrictions on $\epsilon$ are taken into account.
To see this, start on one
side of the configuration, say the left side, with vanishing 10-form.
Let this be the $(+)$ side of the first brane. Then, since
$M_+=0$ we have $M_-=2\mu$ (if we use a brane rather than an anti-brane)
so $Y(R)=\mu^2$ and hence
$\epsilon_-=+1$ so $r_-$ increases towards the brane on the $(-)$
side. This is the $(+)$ side of the next brane in the
configuration. So for this next brane, $M_+=2\mu$ and $\epsilon_+
= +1$ (since $r_+$ decreases towards this new brane). If we now
want $M_-=0$ (i.e. vanishing 10-form) on the other side of the brane then
$[M] = 2\mu$ so it would actually be an anti-brane. 
It would have $Y(R)=-\mu^2$,
which requires $\epsilon_+=-1$, a contradiction. (Intuitively this is
because we are trying to construct a static brane-antibrane
configuration with non-vanishing field strength between them.)
Therefore the second brane in the configuration could not be an
anti-brane. Hence it must have $M_-=4\mu$, and the argument
repeats. One obtains stack of branes with the 10-form field strength
increasing as each brane as crossed. Furthermore $R$ decreases as each
brane is crossed so one must eventually reach the singularity $R=0$ in
finite proper distance.

Note that is \emph{is} possible to construct a static brane-antibrane
configuration but the field strength has to vanish in the region
between the brane and anti-brane. So one can have a brane and anti-brane
separated by a region of Minkowski space, with the ten-form
non-vanishing (and a timelike singularity) on both sides of the configuration.

\subsection{The case $[m]=0$.}

The equation of motion for the brane simplifies when $m_+=m_-=m$. This
requires $m>0$. Integrating gives
\be
 R(\tau) = \left(17\sqrt{2m}|\tau|\right)^{1/17},
\ee
and hence
\be
 r_{\pm}(\tau) = \left(17\sqrt{2m}|\tau|\right)^{26/17}.
\ee
The solutions for positive and negative $\tau$ are related by
time reversal so we shall only consider the expanding (positive
$\tau$) ones. The maximal bulk spacetime has
cosmological horizons beyond which $r$ becomes a time coordinate. At
large $r$, i.e. late times, the metric can be written
\be
\label{eqn:latetime}
 ds^2 \sim -d\tau^2 + \tau^{18/17} dx^2 +
\left(17\sqrt{2m}|\tau|\right)^{2/17} d\Omega_0^2,
\ee
where $x$ is a rescaling of $t$ and we have replaced $r$ with $\tau$. 
The brane sits at constant $x$ and 
is comoving with the spatial sections of this anisotropic cosmological
solution. The direction transverse to the brane expands faster than
directions tangential to it.

\begin{figure}
\centerline{\epsfxsize=15cm \epsfbox{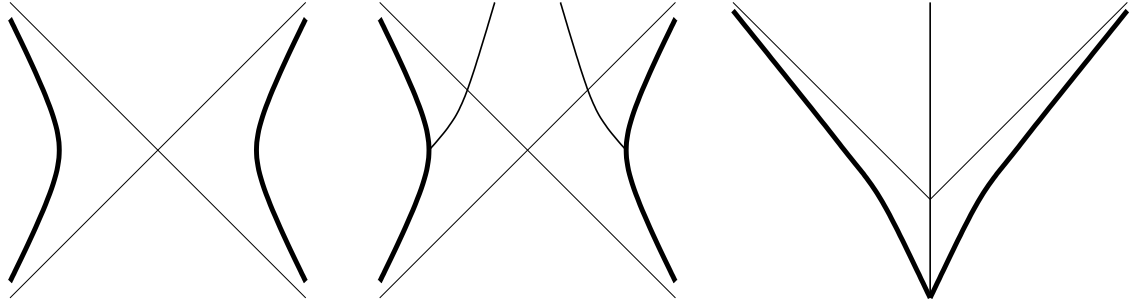}}
\caption{i) The global structure of the maximally extended bulk
spacetime. The thick lines are timelike singularities and the thin
lines are cosmological horizons. ii) The possible domain wall trajectories. 
iii) Global structure of the domain wall spacetime.}
\label{fig:global}
\end{figure}

Since the RR field is produced
by D8 branes, it seems reasonable to assume that $M_\pm$ are integer
multiples of the brane charge $\mu$. 
If we assume that $M_-$ is an integer multiple of $\mu$ then the
solution $\epsilon_+=-1$, $\epsilon_-=+1$ is only possible
for $M_-=\pm \mu$  and hence $M_+=\mp \mu$ 
(as usual the upper sign refers to a brane and the
lower sign to an anti-brane). This means that bulk solution is
the same on each side of the brane and is symmetric under reflection
in the brane. (All of the solutions in \cite{great} are of this form.)
The global structure of the maximally extended
bulk solution is shown in figure \ref{fig:global}. 
There are two possibilities for the brane
trajectory. Let $X_1$ denote the region between such a trajectory and
the singularity from which it emerges, and $X_2$ the complement of
this region. To deduce which
part of this is relevant when we include the effect of the brane, note
that the signs of $\epsilon_{\pm}$ require that $R$ increase towards the
brane along its normal. Thus $X_1$ is the relevant region. 
Cutting along the brane and
gluing to the mirror image of $X_1$ yields the spacetime produced by the
brane. It includes the timelike naked singularity from which the brane
emerges. Initially signals from the brane can reach the singularity
but the presence of a horizon prevents this at late times.
 
\bigskip

In the non reflection symmetric case we can write $M_+ = \pm \mu(N-1)$
and $M_- = \pm \mu(N+1)$ where $N$ is an integer. Consider
first the case $|N|>1$. The solution has $\epsilon_+ = \epsilon_- =
\textrm{sign}(N)$ therefore $R$ must decrease towards the brane on one side and
increase towards it on the other so we need a region of type $X_2$ on one
side of the brane and $X_1$ on the other, although the two regions
have different values for $M^2$ and therefore the brane crosses the
horizon in the region with the smaller value of $M^2$ before it
crosses the horizon in the other region.

The cases $N=1,-1$ are of interest since then the 10 form vanishes on
one side of the brane i.e. the bulk on that side is a solution of
massless IIA supergravity. In this special case the cosmological
horizon vanishes and $r$ is a time coordinate throughout the massless
region. Equation \ref{eqn:latetime} then holds exactly throughout this
region. There is a spacelike singularity at $\tau=0$. 
The region on the other side of the brane is of type $X_1$.

\subsection{The case $[m]\ne 0$.}

When $[m]\ne 0$ it is no longer possible to find $R(\tau)$
explicitly. We shall assume  $m_+>m_->0$.

\begin{figure}
\centerline{\epsfxsize=7cm \epsfbox{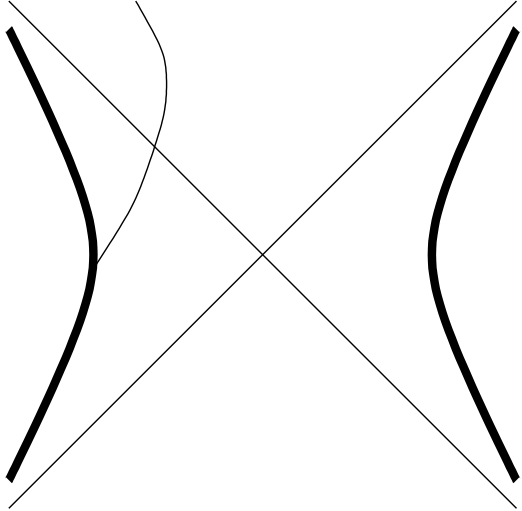}}
\caption{Trajectory of brane with turning point.}
\label{fig:global2}
\end{figure}

Consider first the case $M_+ = \mp \mu$, $M_-=\pm \mu$.
For small $R$, the solution is similar to the
$[m]=0$ one. However for larger $R$, 
\be
 R(\tau) \approx \left(\frac{72}{\mu}[m]\tau\right)^{1/9}.
\ee
At late times, the bulk solution on each side of the brane is of the
cosmological form given above i.e. the 8 dimensional 
spatial sections in the bulk perpendicular
have scale factor $\tau^{1/17}$. Therefore the
brane expands faster than the bulk in these directions. This is
because the velocity of the brane in the (spatial) $t$-direction does not tend
to zero at large $R$ as it did for the $[m]=0$ case.
For small $R$, $\epsilon_+ = -1$ and $\epsilon_-=+1$ but at the
critical value
\be
 R^{16} = \frac{\mu^2}{2^7[m]},
\ee
$\epsilon_+$ changes sign. This occurs in the region where $t_+$ is a
spatial coordinate and corresponds to the brane reaching a turning
point in the $t_+$-coordinate i.e. changing direction of motion relative
to the spatial sections of the bulk spacetime on the $(+)$ side. There
is no such turning point on the $(-)$ side. The trajectory of
the brane spacetime with respect to the maximal bulk spacetime
corresponding to the $(+)$ side of the brane is 
shown in figure \ref{fig:global2}.
Considering the signs of $\epsilon_{\pm}$ near $R=0$ shows that it is
the region between the brane and the singularity from which it emerges
that is relevant on both sides of the brane. 
This means that the global structure of the resulting brane spacetime
is very similar to the $[m]=0$ one.

If we consider a brane with vanishing 10-form on the $(+)$ side
then we have $M_+=0$ and $M_-=2\mu$ so
\be
 F(R) = -\frac{32}{\mu^2}[m]^2R^{-16}- m_+ R^{-32}.
\ee
The solution for $R(\tau)$ is qualitatively the same as the solution
we have just discussed. However we now have $Y(R)>\mu^2$ so
$\epsilon_+ = \epsilon_- = +1$. On the $(+)$ side of the brane, the
solution is of the cosmological form discussed above and the brane
emerges from a spacelike singularity. On the $(-)$ side the relevant
region of the maximal bulk spacetime is the region between the brane
and the timelike singularity from which it emerges. 
 
Note that for certain choices of the parameters, $F(R)$ is positive for
small $R$. This prevents the brane from either emerging from
or colliding with the singularity.
For such cases, the solutions describe branes that
collapse down from infinite $R$ to non-zero $R$ and then
re-expand. However the most natural choices for $M$ for a single brane
are the symmetric
choice $M_+=-M_-$ and the choice for which the 10-form vanishes on one
side of the brane. For these choices $F(R)$ is negative everywhere.
Positive $F(R)$ presumably corresponds to there
being a background 10-form field present. Since this would have to be
produced by other D8 branes it would lead us to dynamic
multi-brane configurations, which we shall not discuss here.  

\section{The theory of Howe, Lambert and West}

\label{sec:hlw}

In the Romans theory the mass 
arises from a Higgs mechanism whereby the two-form ``eats'' the vector.
However, as Howe, Lambert and West (HLW) \cite{neil} have recently pointed 
out there are actually three
distinct Higgs mechanisms, corresponding to whether the vector eats the scalar,
the two-form eats the vector or the three-form eats the two-form.  
In this way, they have
constructed a {\it new} massive IIA supergravity, which has the appealing
property that it can be obtained by compactification of 
eleven-dimensional Minkowski
space on a circle, with the introduction of a Wilson line.
In this paper, we will always be assuming that the four-form field strength
(coming from the eleven dimensional supergravity bosonic field sector)
is turned
off.  With this caveat, we then obtain the below equations of motion in
ten dimensions for the HLW theory:
\begin{eqnarray}
R_{ab} - \frac{1}{2}g_{ab}R &=& -2(D_{a}D_{b}{\phi} - g_{ab}D^{2}{\phi} + 
g_{ab}(D\phi)^{2}) \nonumber \\
&  & + \frac{1}{2}(F^{ac}{F_{b}}^c - \frac{1}{4}g_{ab}F^2)e^{2\phi} -
18m(D_{(a}A_{b)} -g_{ab}D^{c}A_{c}) \nonumber \\
&  & - 36m^{2}(A_{a}A_{b} + 4g_{ab}A^2) - 12mA_{(a}{\partial}_{b)}{\phi} \nonumber \\
&  & - 30mg_{ab}A^{c}{\partial}_{c}{\phi} - 144m^{2}g_{ab}e^{-2\phi} + \textrm{fermions} 
 \>
\end{eqnarray}

\begin{equation}
\label{eqn:maxwell}
D^{b}F_{ab} = 18mA_{b}{F_a}^{b} + 72m^{2}e^{-2\phi}A_{a} -
24me^{-2\phi}{\partial}_{a}{\phi}
\end{equation}
\begin{eqnarray}
6D^{2}{\phi} - 8(D\phi)^2 = -R + \frac{3}{4}e^{2\phi}F^2 + 360m^{2}e^{-2\phi}
+ 288m^{2}A^2 +96mA^{b}{\partial}_{b}{\phi} 
- 36mD^{b}A_b  \>
\end{eqnarray}
where $F_{ab} = {\partial}_{a}A_{b} - {\partial}_{b}A_{a}$ as usual.
The mass parameter $m$ determines the effective cosmological constant 
generated by the ten-form field strength.
In this paper, we will be simplifying things by truncating these equations
of motion by setting
the vector field to zero:
\[
A_{(1)} = 0
\]
With this assumption, things simplify considerably.
Indeed, the only non-trivial remaining equations are
the Einstein equation:
\begin{equation}
\label{eqn:einstein}
R_{ab} = 360m^{2}e^{-2\phi}g_{ab}
\end{equation}
together with the ``Maxwell'' equation \ref{eqn:maxwell}, which implies that
the dilaton $\phi$ is a constant.
In other words, if we turn off all of the fields in this theory
except for gravity, we simply recover de Sitter spacetime.  The effective
cosmological constant is then given explicitly in terms of the mass as
\[
{\Lambda} = 1296m^{2}e^{-2\phi}
\]

\subsection{10-form formulation of the HLW theory}

In \cite{neil} the authors introduce the exact one form
\begin{equation}
k = Mdy
\end{equation}
(In order to avoid annoying numerical factors we have replaced the $m$
of equation \ref{eqn:einstein} with $M$, where $M^2 = m^{2}/630$.)
This vector is both the tangent vector to the eleventh
dimension (the $S^1$) of M-theory, as well as the form used to define
a new derivative $D = d + k$.  It is natural to dualize this one-form
relative to the eleven-dimensional Hodge star operator.  When we do this
we obtain a ten-form $F_{10}$ which is manifestly covariantly constant:
\[
{\star}_{11}F_{10} = k
\]

We can use this 10-form to write down a (truncated) action for the
HLW theory.  Remembering that we are always working only with the sector
of the HLW theory where the dilaton is constant, we find that the 
string frame action takes the form
\begin{equation}
 S = \frac{1}{2\kappa_{10}^2} \int d^{10}x \sqrt{-g} \left(e^{-2\phi} (R +
4(\partial \phi)^2) -\frac{1}{2}e^{-4\phi} {F_{10}}^2 \right)
\end{equation}

Clearly, the first thing one notices about this action is the strange
factor of $e^{-4\phi}$ which appears in front of the 10-form.  
This means that the 10-form of the HLW theory is not a R-R sector field
and consequently the HLW 8-branes are not D-branes.  In fact, this action
teaches us something even stranger: If we match dimensions in (5.5)
we find that the tension (T) of a brane which couples electrically 
to $F_{10}$ must scale as
\[
T \sim e^{\phi}
\]

In other words, in the HLW theory the 8-branes have a tension that scales with the 
`string coupling'!  \footnote{It is worth pointing out here, that since 
the 3-form has eaten the 2-form, there is no field to which F-strings
can couple electrically.  In this sense, there are no `strings' in 
this theory!}  Since we are interested in the gravitational effects of these
objects, we are implicitly assuming that $\phi$ is very large in what follows.

Our plan now is simple: The eight-brane divides spacetime up into
separate domains, and we use the Israel conditions to match 
these domains correctly. We now turn our attention to this exercise. 

\subsection{Spherical phase transition bubbles in the HLW theory}

By assumption, we are working in the regime of the HLW theory where the
dilaton is constant.  Consequently, each side of a HLW 8-brane is a portion
of a spacetime which is a solution of the vacuum Einstein equations with
positive cosmological constant in ten dimensions.
In other words, each side of the brane is a portion of de Sitter space,
or Schwarzschild-de Sitter space.  In general, we can find timelike slices
of (Schwarzschild)-de Sitter spacetime which may correspond to the worldvolumes
of homogeneous and isotropic 8-branes with arbitrary spatial
curvature. This is unlike the case of D8 branes studied above, for
which we could only find solutions with Ricci flat spatial
sections. We will focus on branes with spherical
symmetry\footnote
{The general case was studied in an earlier version 
of this paper \cite{old}, where we found all of the homogeneous and
isotropic 8-branes of the HLW theory.}. 

To do this we simply choose a gauge so that the spherically
symmetric ten-dimensional metric takes the form \cite{tang}, \cite{rob}:
\begin{equation}
ds^2 = -f(r)dt^2 + \frac{1}{f(r)}dr^2 + r^{2}d{{\Omega}_{8}}^2
\end{equation}
The most general spherically
symmetric {\it vacuum} metric in $9 + 1$ dimensions will have
\begin{equation}
\label{eqn:metf}
f(r) = 1 - \frac{C}{r^7} - \frac{\Lambda}{36}r^2.
\end{equation}
This is of course just the metric of Schwarzschild-de Sitter (SdS) spacetime,
where $\Lambda$ is the cosmological constant and $C$ is given in terms
of the ADM mass $M$ mass of the black hole as follows:
\[
C = (\frac{105G}{16{\pi}^3})M
\]
Actually, we could also leave $F_{{\mu}{\nu}}$ non-zero and 
consider the resulting Einstein-Maxwell-dilaton system of equations; we
begin with the $F = 0$ truncation for simplicity.  

With this simple picture in mind we can now proceed with the equations of motion.
The equations of motion for a thin spherical shell, which is bounding a region 
of SdS from a region of a different SdS
spacetime, were worked out for four dimensions in \cite{ber}.  The analysis
developed there will go through here, because the assumption
of spherical symmetry means that brane motion is described as a system in
$1 + 1$ dimensions.  Put more simply, we are solving for one unknown function,
$r(t)$, which describes the radius of the spherical brane at time $t$.
We will let $\sigma$ denote the surface energy
density of the brane.  We will also distinguish between quantities which
exist ``inside'' of the brane, and those which are outside, by using 
subscripts.
Thus, $f_{in}$ denotes the metric function \ref{eqn:metf} inside the brane,
and similarly $f_{out}$ denotes the same thing outside the brane.
The parameters $\epsilon$ have the same interpretation as they had for
our D8-brane solutions provided we take the unit normal to point {\it
out} of the bubble i.e. the interior of the bubble is the $(+)$
region. With all of this in mind, the equation of motion for a spherical 
bubble bounding two different 
(vacuum) phases of the HLW supergravity theory is given as
\begin{equation}
\label{eqn:hlwmove}
{\epsilon}_{in}[{\dot r}^2 + f_{in}(r)]^{1/2} - 
{\epsilon}_{out}[{\dot r}^2 + f_{out}(r)]^{1/2} = -\frac{{\kappa}\sigma}{2}r
\end{equation}
where ${\dot r} = f^{-1/2}{\partial}_{t}r$ denotes the derivative
relative to the time experienced by observers who co-move with the domain wall
and ${\kappa} = 8{\pi}G$ as always.
Any solution of this equation automatically describes an eight-brane worldsheet
which satisfies the Israel conditions.

Again, it is easy to see that one can generically rearrange the terms in
\ref{eqn:hlwmove} so that the equation of motion for 
the scale factor is equivalent to 
the equation of motion for a particle of unit mass and zero total energy moving
in a potential $V$.  The precise form for $V$ will depend on what one chooses
to lie on each side of the 8-brane.  Because this is a straightforward 
generalization of work which has already been done in the 
setting of four-dimensional cosmology,
we will have nothing more to say about these branes here.

\subsection{Semiclassical instabilities of HLW 8-branes}

The nucleation and annihilation of bubbles of phase transition in the early
universe has been studied by a number of authors \cite{shawn1}, \cite{guth},
\cite{ber}, \cite{fgg}.  Here, we briefly sketch how these results will
also go through for HLW branes.

In \cite{fgg}, the authors calculated the probability that one could
``create a universe in the laboratory'' by quantum tunneling.  By this, they
meant: What is the probability that a bubble of false vacuum (i.e.,
de Sitter space)  can appear in a lab where we have attained a super-high
mass density of the order $10^{76}$ g/$cm^{3}$?  
The authors calculated the rate at which one could create new
universes in this way by first finding an instanton, or imaginary time path, which
interpolates between the initial state (Schwarzschild) and the final state
(Schwarzschild with a bubble of de Sitter in it), then working out the
Euclidean action $S_E$ for the instanton path and then using the standard
semiclassical approximation for the probability $P$:
\[
P {\propto} e^{-S_{E}}
\]

\noindent Crudely, they found that the Euclidean action goes like
\[
S_E {\propto} 1/H^2
\]
\noindent where $H = (\frac{\Lambda}{3})^{1/2}$ is the inverse of the Hubble
radius.  This form of the action follows from the fact that the domain wall
sweeps out a three-sphere in imaginary time, and the action is basically the
wall tension times the volume of the three-sphere.

In ten dimensions, an HLW 8-brane sweeps out a nine-sphere in imaginary
time; it follows that the Euclidean action for the nucleation of HLW branes
will go like $1/H^8$.  In other words, the nucleation of isolated
bubbles of massive
phase in this IIA supergravity theory will still be highly suppressed.

This shows that HLW branes can be spontaneously nucleated, but what about
the time reverse: How do they annihilate?  In a recent paper,
Kolitch and Eardley \cite{shawn1} studied the decay of  
Vilenkin-Ipser-Sikivie (VIS) \cite{vis} 
domain walls in cosmology.  These VIS domain
walls are the Minkowski-Minkowski version of the de Sitter-de Sitter (dS-dS)
HLW branes discussed above.  That is to say, a VIS domain wall is a repulsive
spherical bubble which separates two (compact) portions of Minkowski space
(just as a dS-dS HLW brane separates two compact portions
of de Sitter spacetime).  It is not hard to see that their construction
will in fact go through for the dS-dS HLW branes.  Again, the action for 
the instanton describing such an annihilation event will go crudely as
$1/H^8$.  

It is clear that black hole pairs will be nucleated in the
presence of the repulsive, dS-dS spherical HLW branes.  This is basically
because these dS-dS branes are bubbles bounding two regions of inflationary
phase, and we know that black holes will be produced in an inflating
(or domain wall \cite{pairs}) background. 
Of course, just about {\it anything} can
be produced in an inflationary background, simply because
the repulsive gravitational energy is a natural source capable
of pulling virtual loops of matter out of the vacuum.
In particular, it is well known that topological defects \cite{review}
will also be nucleated in inflation.
Typically, a defect nucleates at the Hubble radius $r = H^{-1}$.
If the defect is much thinner than the scale
of the universe at the moment of nucleation, it makes sense to model the
defect using the Nambu action.  That is, it makes sense to assume that the 
defect is ``infinitely thin'', and to define the action to be the area of the
worldvolume swept out by the defect (multiplied by the characteristic tension,
or mass, of the defect).  In this limit, the instanton for a loop of 
string is a two-sphere of radius $H^{-1}$.  Similarly, the instanton for a 
(closed) spherical
domain wall is a three-sphere of radius $H^{-1}$, and so on.
Thus, the Euclidean action
for defect nucleation generically has the form
\[
S_{E} = {\mu}Vol(S^{n}(1/H))
\]
\noindent where $Vol(S^{n}(1/H))$ denotes the volume of an n-sphere of radius
$1/H$, and $\mu$ denotes the mass of the monopoles (if $n = 1$), the tension of
the string loop (if $n = 2$) or the energy density of the domain wall (if $n = 3$). 
One therefore finds that the rate of production of these defects is strongly
suppressed if the defect tension is very large, or the cosmological constant
is very small, as would be expected.  Similarly, light defects are likely to
be produced in a background with a large cosmological constant.

\section{Conclusion}

\label{sec:conclude}

We have found solutions of the coupled bulk-worldvolume theory
which describe the motion of a D8-brane in massive IIA supergravity.
Since the dilaton is running in any such background,
it follows that the energy-momentum carried by that field is allowed
to move from the brane to the bulk and vice-versa.  We have also shown
that whenever a D8-brane starts to move, a non-extremality term 
will necessarily appear in the bulk on both sides of the brane. This term
obstructs the existence of
Killing spinors and therefore breaks any supersymmetry which was initially
present. The spacetime on either side of the brane at late times is
typically an anisotropic cosmological solution with the direction
transverse to the brane expanding faster than the directions
tangential to it. When the non-extremality parameter is the same on both
sides of the brane the brane eventually becomes comoving with the
bulk. If the non-extremality parameters are not the same then the
brane always moves relative to the bulk.

Use of the Israel conditions also places constraints on supersymmetric
solutions. We have shown that it is not possible to exclude the
singularity from a supersymmetric D8-brane spacetime (unless one
introduces objects of negative tension, such as orientifold planes). 
We have also
shown that it is impossible to construct a configuration of parallel
D8 branes (or anti-branes) that is a solution of the \emph{massless} IIA
theory in the two asymptotic regions. 

We have also discussed branes in the Howe-Lambert-West theory. Unlike
our D8 brane solutions, these can have spatial sections with Ricci
curvature, and such spherical branes might be relevant for the study
of phase transitions in the supergravity theory. It would be
interesting to see if spherical D8 brane solutions could be found,
although this would probably have to be done numerically. If such
solutions exist then they could describe the evolution of a bubble of
the massive phase of IIA supergravity surrounded by the massless phase.

Perhaps the most exciting thing which these results teach us is that
it is possible to describe gravitating brane configurations without losing
sight of the brane worldvolume.  To put it another way, there is nothing to
stop one from defining an effective worldvolume action for these gravitating
D8-branes (presumably the Born-Infeld-Dirac action will do), and studying
these branes from the worldvolume point of view.  Again, since the dilaton
will flow into the bulk, 
there will be an anomaly in the stress-energy conservation
equation for the worldvolume theory.  It would be amusing to see how this is
related to the breakdown of supersymmetry.

This should be contrasted to other heavy, 
or gravitating, brane configurations in supergravity
theories, such as the non-dilatonic p-branes \cite{gazpaul}.  There, the
brane worldvolume vanishes, leaving a geometry which looks
rather like a black hole (extended in some extra dimensions), such that the
solution interpolates between different vacua (generically Minkowski 
spacetime far from the brane, and $(adS)_{p+2} {\times} S^{D - p -2}$
near the brane \cite{gazpaul}).  
In these solutions, there is no brane to be found and 
it is therefore meaningless to talk about worldvolume actions.

The results \cite{gazz,curt} concerning
how a string ending on a D-brane will tug on the brane beg the 
question:  How will a heavy string tug on a heavy D8-brane?  One has
to be careful in posing this question, since it is not at all clear that
the method of assigning distributional curvatures to spacetimes of low
differentiability (i.e. imposing the Israel conditions) 
will still make sense when one
is dealing with extended objects of codimension greater than one. 
Research into these problems is currently underway.

\bigskip

\centerline {\bf Acknowledgements}

The authors would like to thank Gary Gibbons, Mike Green, Julius Kuti,
and Neil Lambert for useful
conversations.  A.C. is supported by a Drapers Research Fellowship at
Pembroke College, Cambridge.  M.J.P. was partially supported by a grant
from the U.S. Department of Energy.


\begin{thebibliography}{99}

\bibitem{joe} J. Polchinski, {\it Dirichlet-Branes and Ramond-Ramond
Charges}, Phys. Rev. Lett. {\bf 75}, 4724 (1995); hep-th/9510017.

\bibitem{dai} J. Dai, R.G. Leigh and J. Polchinski, {\it New Connections
between String Theories}, Mod. Phys. Lett. {\bf A4}, 2073 (1989).

\bibitem{larus} L. Thorlacius, {\it Introduction to D-branes},
Nucl. Phys. Proc. Suppl. {\bf 61A}: 86-98, (1998);
hep-th/9708078.

\bibitem{gazz} G.W. Gibbons, {\it Born-Infeld particles and Dirichlet
p-branes}, Nucl. Phys. {\bf B514}: 603-639, (1998); hep-th/9709027.

\bibitem{curt} Curtis G. Callan, Jr. and J.M. Maldacena, {\it Brane death
and dynamics from the Born-Infeld action}, 
Nucl. Phys. {\bf B513}: 198-212, (1998); hep-th/9708147.

\bibitem{great} A. Chamblin and H.S. Reall, {\it Dynamic Dilatonic
Domain Walls}, hep-th/9903225, submitted to Nucl. Phys. {\bf B}.

\bibitem{ber} V.A. Berezin, V.A. Kuzmin, and I.I. Tkachev, {\it Dynamics
of bubbles in general relativity}, Phys. Rev. D {\bf 36}, No. 10,
2919-2944 (1987).

\bibitem{guth} Steven K. Blau, E.I. Guendelman, and Alan H. Guth,
{\it Dynamics of false-vacuum bubbles}, Phys. Rev. D {\bf 35}, No. 6,
1747-1766 (1987).

\bibitem{israel} W. Israel, Nuovo Cimento {\bf 44B}, 1 (1966); {\bf 48B},
463E (1967).

\bibitem{cvetic} M. Cvetic, S. Griffies and H.H. Soleng, {\it Local and global
gravitational aspects of domain wall space-times}, Phys. Rev. D {\bf 48},
2613-2634 (1993); gr-qc/9306005.

\bibitem{cvetic2} N. Cvetic and H.H. Soleng, {\it Naked singularities
in dilatonic domain wall space-times}, Phys. Rev. {\bf D51}, 5768-5784
(1995); hep-th/9411170.

\bibitem{mir} M. Cvetic and H.H. Soleng, {\it Supergravity domain walls},
Physics Reports 282, 159-223 (1997), hep-th/9604090.

\bibitem{roman} L.J. Romans, {\it Massive $N = 2a$ supergravity in ten
dimensions}, Phys. Lett. {\bf 169B}, No. 4, 374-380 (1986).

\bibitem{bergshoeff} E. Bergshoeff, M. de Roo, M.B. Green,
G. Papadopoulos and P.K. Townsend, {\it Duality of Type II 7-branes
and 8-branes}, Nucl. Phys. {\bf B470}, 113-135 (1996); hep-th/9601150.

\bibitem{dom} D. Brecher and M.J. Perry, {\it Ricci-Flat Branes}, 
hep-th/9908018.

\bibitem{neil} P.S. Howe, N.D. Lambert and P.C. West, 
{\it A new massive Type IIA
supergravity from compactification}, Phys. Lett. {\bf B416}: 303-308, (1998);
hep-th/9707139 v2.

\bibitem{old} A. Chamblin and M.J. Perry, hep-th/9712112.

\bibitem{rob} R.C. Myers and M.J. Perry, {\it Black holes in higher dimensional
space-times}, Annals of Physics {\bf 172}, 304-347 (1986).

\bibitem{tang} F.R. Tangherlini, {\it Schwarzschild field in n dimensions
and the dimensionality of space problem}, Il Nuovo Cimento, Vol. XXVII,
636-651, (1963).

\bibitem{shawn1} Shawn J. Kolitch and Douglas M. Eardley, {\it Quantum
decay of domain walls in cosmology: 1. Instanton approach},
Phys. Rev. D {\bf 56}: 4651-4662, (1997);
gr-qc/9706011.
 
\bibitem{vis} A. Vilenkin, Phys. Lett. {\bf B133}, 177-179 (1983);
J. Ipser and P. Sikivie, Phys. Rev. D {\bf 30}, 712-719 (1984).

\bibitem{fgg} E. Farhi, A.H. Guth and J. Guven, Nucl. Phys. {\bf B339}, 417 (1990).
 
\bibitem{pairs} R. Caldwell, A. Chamblin and G.W. Gibbons, 
{\it Pair creation of
black holes by domain walls}, Phys. Rev. D {\bf 53}, 7103-7114 (1996),
hep-th/9602126; A. Chamblin and J.M.A. Ashbourn-Chamblin, {\it Black hole
pairs and supergravity domain walls}, Phys. Rev. D {\bf 57}: 3529-3536, (1998);
hep-th/9612014.

\bibitem{review}  A. Vilenkin and E.P.S. Shellard, {\it Cosmic
Strings and Other Topological Defects}, Cambridge University Press,
Cambridge (1994).


\bibitem{gazpaul} G.W. Gibbons, P.K. Townsend, {\it Vacuum interpolation
in supergravity via super p-branes}, Phys. Rev. Lett. {\bf 71}, 3754-3757, (1993);
hep-th/9307049.

\end{thebibliography}
\end{document}